\pdfoutput=1 % only if pdf/png/jpg images are used
\documentclass{JINST}

\graphicspath{./pictures/}
\usepackage{subfigure}
\usepackage{amsmath}
\usepackage{siunitx}

\DeclareSIUnit[number-unit-product = {}] \bit{bit}
\title{Gbit/s Data Transmission on Carbon Fibres}
\author{T. Flick$^a$\thanks{Corresponding author of University Wuppertal.}, K.-W. Glitza$^a$, G. C. Kappen$^b$\thanks{Corresponding author of Fachhochschule M\"unster}, P. M\"attig$^a$, J. M\"oller$^a$ and B. Sanny$^a$\\
\llap{$^a$}Bergische Universit\"at Wuppertal, \\
	Gau\ss strasse 20, 42097 Wuppertal,Germany\\
\llap{$^b$}Fachhochschule M\"unster,\\
	Stegerwaldstra\ss e 39, 48565 Steinfurt, Germany\\
E-mail: \email{flick@uni-wuppertal.de}\\
E-mail: \email{goetz.kappen@fh-muenster.de}
}

\abstract{
	Data transmission at the upgraded Large Hadron Collider experiments, foreseen for mid 
2020s will be in the multi \SI[per-mode=symbol]{}{\giga\bit\per\second} range per connection for the innermost detector layers. 
This paper reports on first tests on the possible use of carbon cables for electrical data transmission close to the interaction point. Carbon cables have the potential advantage of being light, having a low activation and easy integration into the detector components close to the interaction point. In these tests commercially available carbon cables were used, in which the filaments had a very thin nickel coating. For these cables data rates beyond \SI[per-mode=symbol]{1}{\giga\bit\per\second} over more than \SI{1}{\meter} with an error rate of less than \num{e-12} could be transmitted. The characteristics of the cables have been measured in terms of S-parameters and could be reproduced with a Spice model. Some outlook on potential further improvements is presented.

}


\begin{document}
	
\section{Introduction}
\label{sec:intro}
The upgraded Large Hadron Collider (LHC)~\cite{bib-HiLumLHC}, planned for mid 2020s will deliver unprecedented high data rates. By colliding protons with highest energy and intensity the LHC will allow the experiments to explore in depth building blocks of matter in distances smaller than \SI{e-19}{\meter} . The upgraded LHC will explore and possibly reach beyond the limits of the Standard Model of particle physics. Whereas, on the one hand, data will be collected in a much harsher environment of high particle flux than before, on the other hand the extremely high precision of the measurement with current experiments has to be retained, if not improved. These goals impose challenges for both the LHC accelerator and its detectors.

Especially, the innermost parts of the LHC detectors, which are used to detect the tracks of charged particles, have to meet new levels of performance requirements, demanding new technological solutions. They will be based on radiation hard, finely granular silicon sensors.
While these detectors
have to be ultralight but stable to retain the very high internal precision, they also have to be radiation hard and still be able to transmit a total of several \SI[per-mode=symbol]{}{\peta\bit\per\second} of data.
Furthermore, to allow for replacements and eventual removal, the activation due to the
high radiation level has to be kept as low as possible.

One of the critical elements of such new detectors will be a highly reliable system to read out the collected data and to provide steering commands to the detector. While most of some \SI{80}{\meter} between the silicon sensors and the data processing at computer farms will use optical transmission, data on the first some \SI{5}{\meter} between sensors and opto-electrical-converters (opto-transceivers) will be transmitted electrically. In this paper the potentials of carbon cables for high bandwidth data transmission for pixel detectors at the detector centre will be discussed. 

\begin{figure}[b]
	\begin{center}
		\includegraphics[width=\textwidth]{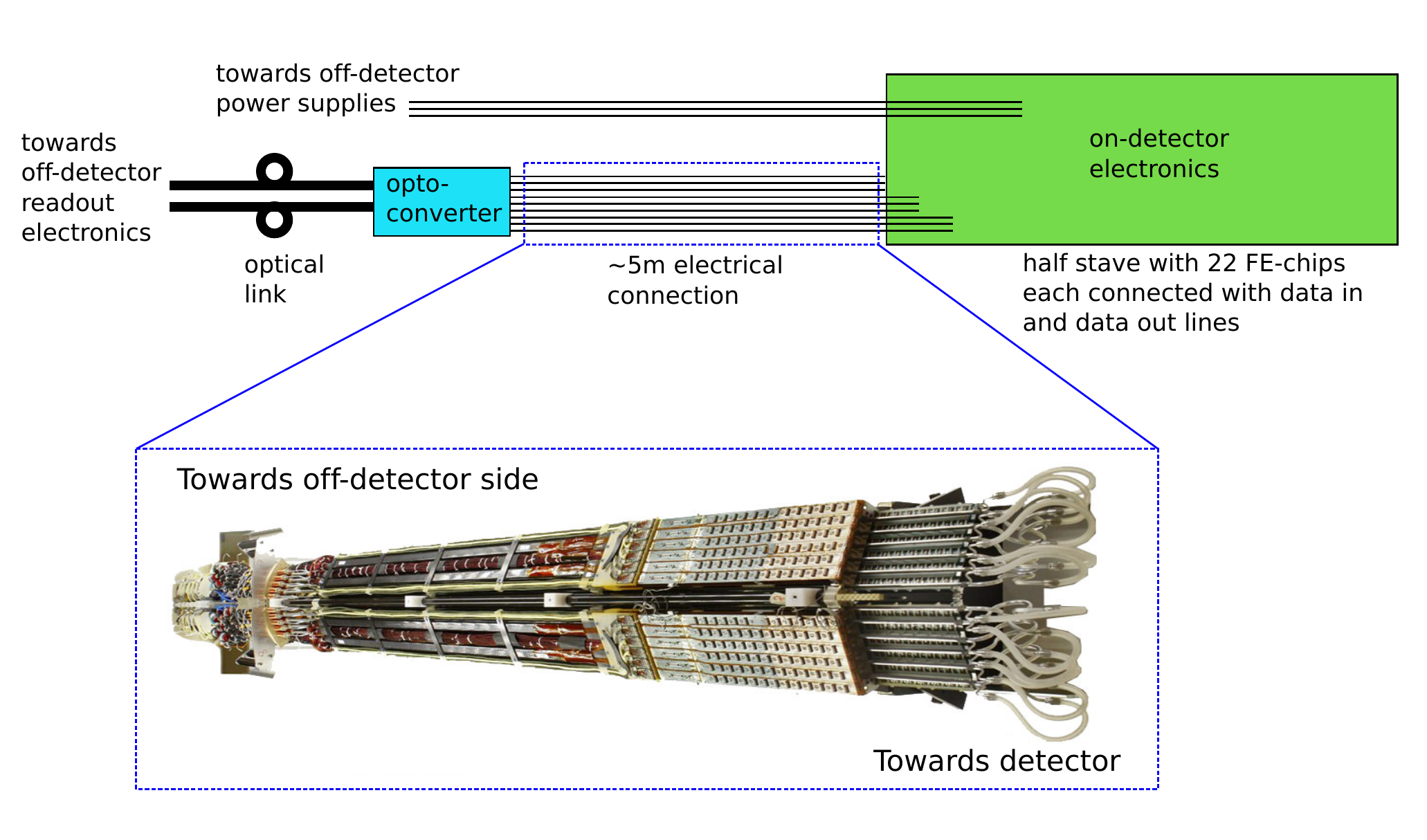}
		\caption{Sketch of read-out and service connections for each half stave of the current ATLAS pixel detector. 
		Per read-out chip there are several data lines needed to transmit data to and from the detector. The innermost part of the cabling is foreseen to be electrically connected to the electrical-to-optical converters. The zoomed-in picture shows a service panel carrying cables and cooling pipes for a quarter of one detector half.}
		\label{fig:readout}
	\end{center}
\end{figure}

To define the environment: as part of the ATLAS detector~\cite{bib-ATLAS} at the LHC, the new pixel
detector will, similar to the current one, will consist of electronically autonomous compartments, the modules, which comprise the silicon sensors and custom made front-end (FE) electronic chips to read out the signal and provide a first processing. These modules are glued on staves, ultralight mechanical carbon structures, which serve to both hold these modules in place and remove the heat generated by electronics and sensors. Data collected in the FE electronics is transmitted electrically along the staves and beyond where it is converted into optical signals. Similarly, steering signals have to be electrically transmitted into the modules. The services to provide electrical connections for supply and read-out to the modules are shown in Figure~\ref{fig:readout}.

%\begin{figure}[h]
%	\begin{center}
%		\includegraphics[width=12cm]{./pictures/readout-scheme}
%		\caption{Schematical view of readout and service connections for each half stave. Per readout chip there are several data lines needed to transmit data to and from the detector. The innermost part of the is foreseen to be electrically feeding the data to the electrical to optical converters.}
%		\label{fig:readout}
%	\end{center}
%\end{figure}
%
%\begin{figure}[h]
%	\begin{center}
%		\includegraphics[width=12cm]{./pictures/nsqp_cables_freigestellt.jpg}
%		\caption{Picture of the current cable structure (including carrier) to support the cables between the detector package and the first patch panel.}
%		\label{fig:nsqp}
%	\end{center}
%\end{figure}

These detectors have to withstand radiation levels of up to \SI{7.7}{\mega\gray} during some 10 years of operation but have to measure the point of a particle transition to  a precision of some \SI{5}{\micro\meter}. Detailed requirements are still under discussion, but it is currently planned that each module will to be read out with up to \SI[per-mode=symbol]{5}{\giga\bit\per\second} on a single transmission line and
the data should be electrically transmitted over a distance of a few meters before converted into optical signals~\cite{bib-ATLAS-LOI}. The signal is DC-balanced and the bit error rate (BER) should be less than \num{e-12}.

Current solutions, which have to cope with much lower radiation doses and bandwidths 
of only \SI[per-mode=symbol]{160}{\mega\bit\per\second}, use copper bands embedded in kapton foils. It is, at this stage, an active R\&D field to arrive at solutions for significantly higher data rates. 
On the one hand copper cables can at least in principle transport electrical signals with high bandwidth, but on the other hand they have disadvantages as to their radiation length and the mismatch of its thermal expansion coefficient (CTE) with respect to the mechanical support. Alternative materials for high frequency electrical transmission would therefore be desirable.

While the potentials of using carbon fibres for electrical data transmission for a high granularity pixel tracker have to be investigated, carbon materials have several potential advantages compared to copper and other metallic materials used for cables:  

\begin{itemize} 
\item Carbon is some factor 3.5 lighter than copper, important to achieve an optimal tracking precision,
\item Carbon has little activation,  
\item Fibres made from carbon have the same CTE as the stave itself, thereby avoiding any additional stress on the stave, even offering the chance to integrate them into the structure.        
\end{itemize}

The key question is if a high data transmission is achievable. In this paper we will discuss first results on the performance of carbon cables, for up to \SI[per-mode=symbol]{1.25}{\giga\bit\per\second}. 
The scope of this paper is to show that \SI{}{\giga\bit} transmission can be reliably achieved with carbon cables. At this stage no optimization has been attempted which will be subject to future R\&D.
Although, the development focuses on the application in an LHC pixel detector, the advantages of carbon could also be used for other kinds of detectors or even outside high energy physics.

\begin{figure}[h]
	\begin{center}
		\includegraphics[width=10cm]{{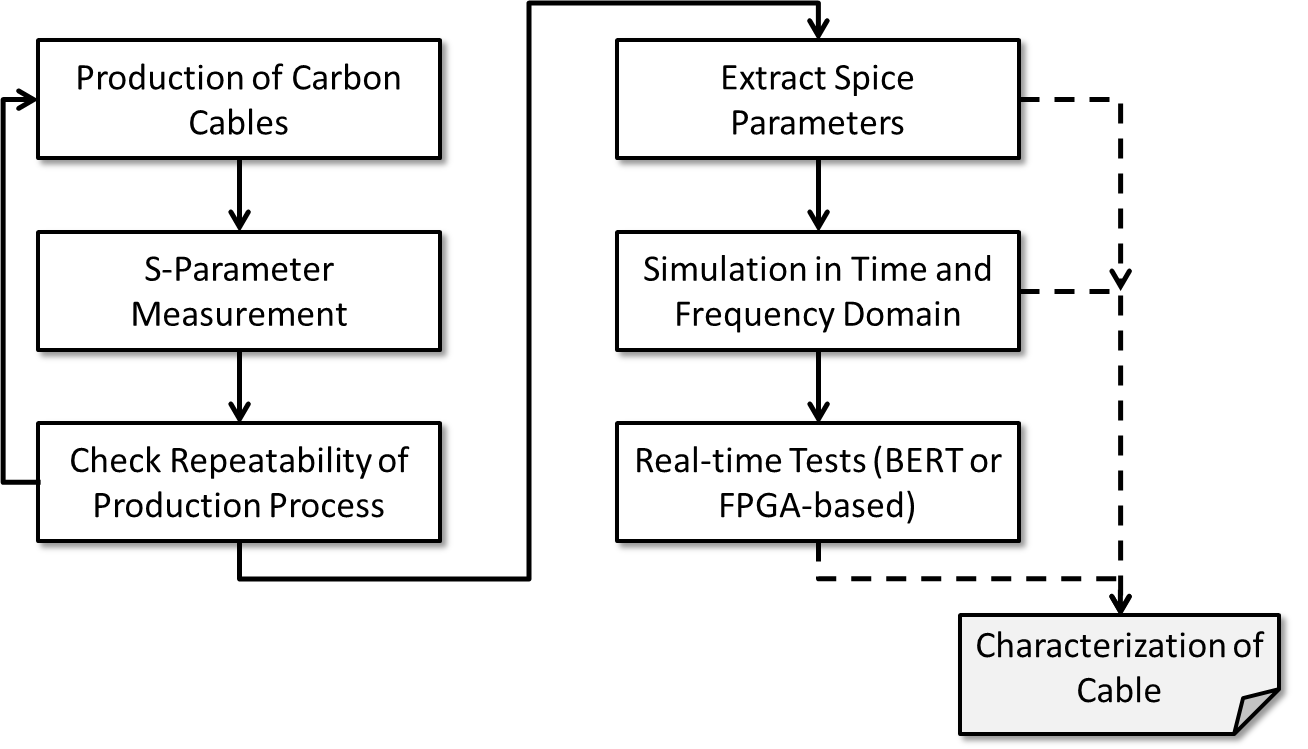}}
		\caption{Proposed cable characterization process}
		\label{fig:prop_design_flow}
	\end{center}
\end{figure}

Overall we follow in the project the design flow shown in Figure~\ref{fig:prop_design_flow} which starts with the production process followed by the measurement and verification of the cable properties. Afterwards, Spice parameters are extracted based on the measured S-parameters allowing simulations of the cable in time and frequency domain. Together with the results of the final real-time BER-test we receive a characterization of the carbon cable transmission line properties in the \SI{}{\giga\bit} range.

\bigskip

The paper follows this flow and
will start by describing the fabrication of carbon cables (Section~\ref{sec:fabrication}), report on first performance (Section~\ref{sec:firsttests}) test and then discuss a model of the cables using mathematical tools and descriptions (Section~\ref{sec:modell}). 
Finally, in Section~\ref{sec:fastfpga} FPGA-based real-time tests realizing \SI[per-mode=symbol]{}{\giga\bit\per\second} data transmission will be presented. 
The paper ends with some conclusions and further directions of work (Section~\ref{sec:Conclusions}).

\section{Fabrication of Carbon Cables}
\label{sec:fabrication}
% \todo[inline, color=yellow]{Produktion, Wiederholbare Herstellung (wichtiger Punkt, hier können S-Parameter Messungen und Analysen untergebracht werden. Außerdem sollte das Kabelmodell hier eingeführt werden [also Kapitel III-A-1)-3)])}

As a starting point of this project, several potential geometries of carbon cables were briefly studied, i.e. coaxial and twisted pair carbon cables. Based of these first results as well as to account for fabrication simplicity and performance the following studies were made with single cables for data transmission. However, to reduce noise a second passive carbon cable was twisted around the signal cable. 

Poly-Acryl-Nitril type fibres were preferred over pitch based fibres to arrive at more reliable cables. Pitch based fibres have lower resistivity and higher tensile strength, but are extremely brittle and difficult to bend. In a first step, without optimising for thickness, HTS40~\footnote{The A23 MC 12K 1420tex MC fibre from Toho Tenax\textsuperscript{\textregistered} Europe GmbH \cite{bib-TohoTenax} was used} fibres were used consisting of \num{12000} filaments of \SI{7}{\micro\meter} thickness each. Each of these filaments was coated with \SI{250}{\nano\meter} of nickel. 
%to improve the performance.
 In total this leads for an individual carbon fibre to a thickness of about \SI{0.9}{\milli\meter}. The carbon cables were inserted into a \SI{0.3}{\milli\meter} thick sleeve of PVC thermoplastic for isolation. In total this leads to a twisted cable of about \SI{2}{\milli\meter} in diameter. 
%,which we can reduce to 1.7mm by using thinner sleeves and this will also reduce the radiation lengths. 

% \subsection{Assembly of CF-cabels}

\begin{figure}[!ht]
	\centering
	\subfigure[][]{\label{fig:kabelhuelse}\includegraphics[width=7cm]{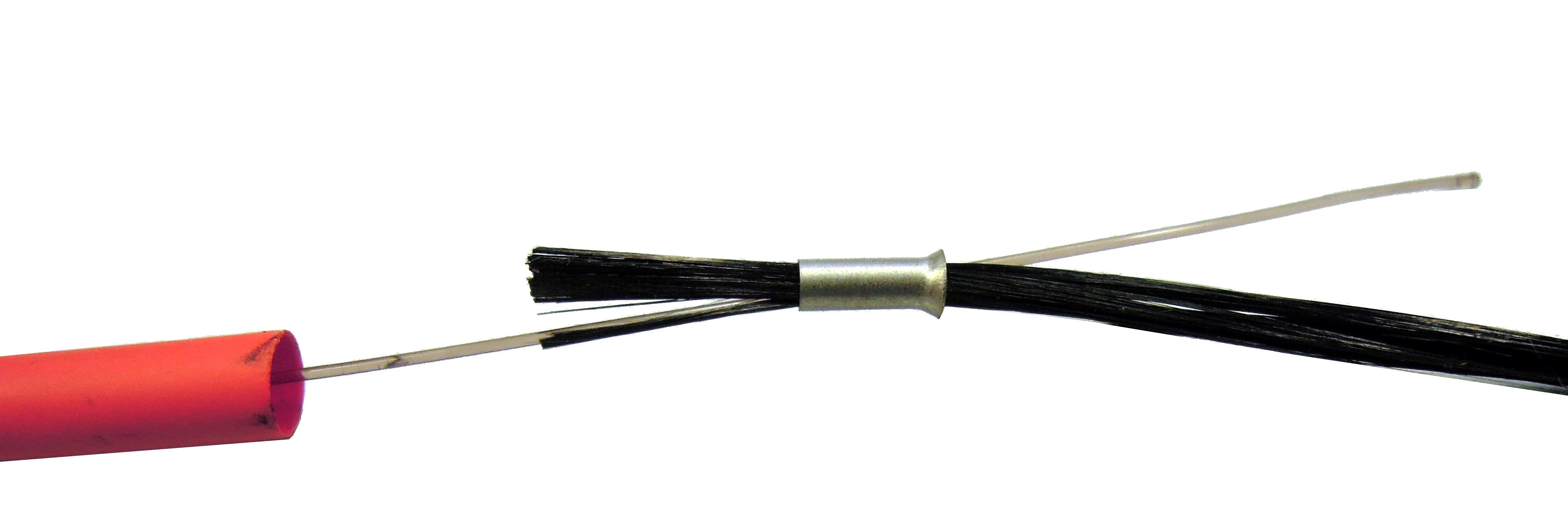}}
	\hspace{0.2cm}
	\subfigure[][]{\label{fig:durchgezogen}\includegraphics[width=7cm]{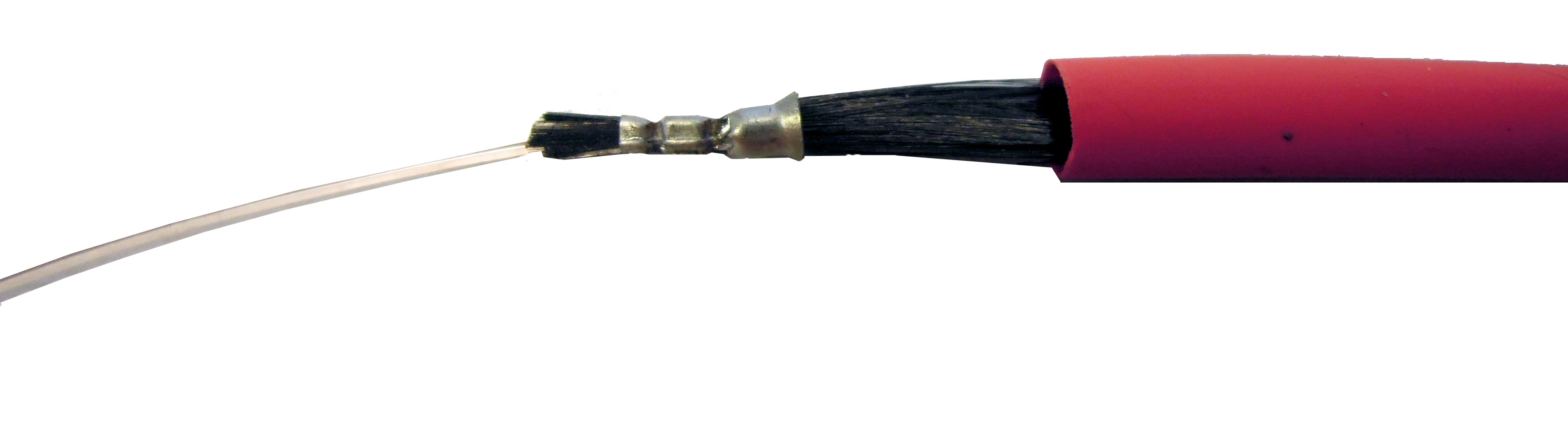}}
	\hspace{0.5cm}
	\subfigure[][]{\label{fig:crimp}\includegraphics[width=7cm]{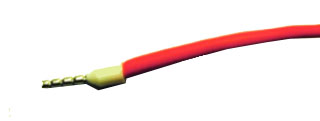}}
	\hspace{0.2cm}
	\subfigure[][]{\label{fig:sma-connector}\includegraphics[width=7cm]{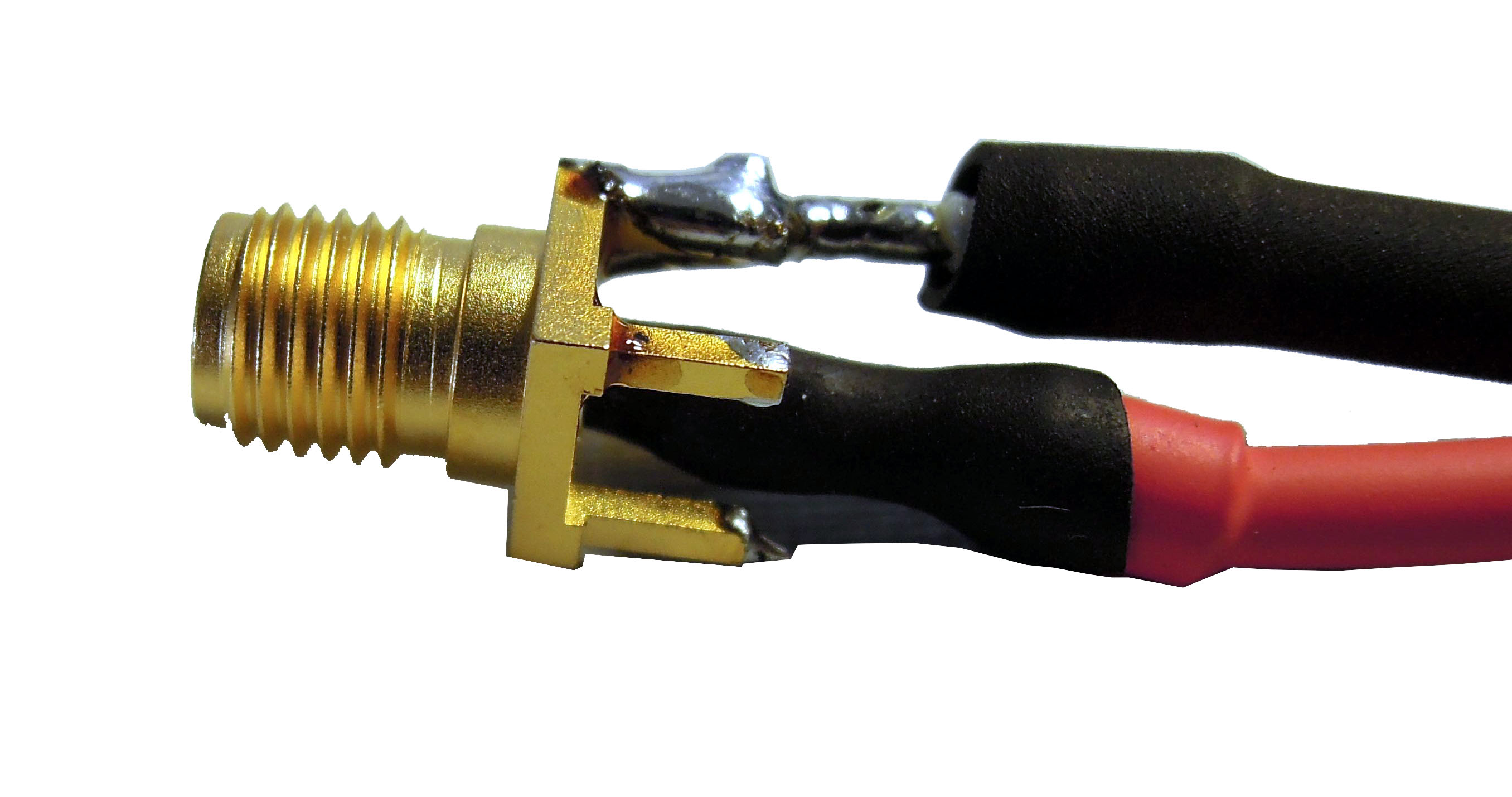}}
	\caption{Major steps of cable fabrication process with crimped terminations soldered to an SMA connector. First the fibres need to be pulled through an PVC sleeve for isolation. For this a pull-wire is connected to the fibres (a), with this the fibres can be pulled into the sleeve (b). A crimp connection terminates the fibres (c) which is then soldered to a connector (d).}
	\label{fig:fabrication}
\end{figure}

To fabricate the cable, carbon fibres of some lengths were cut from a spool and were inserted into a sleeve, with the help of a nylon cable that is steered through the tube and crimped to the fibre bundle (see Figure~\ref{fig:kabelhuelse}). 
This allowed pulling the bundle through the sleeve (see Figure~\ref{fig:durchgezogen}). 
Care has to be taken to not bend the fibres too strongly to avoid broken or disabled filaments. In a final step the sleeve  is shrunk to a third of its original diameter using a hot air gun. 

To check if the cable is unbroken and functioning, its resistance is measured. 
Once correct operation is shown, the cable is cut to the desired length and both ends are crimped (see Figure~\ref{fig:crimp}) allowing an exact connection of the cable by soldering it to an SMA connector (see Figure~\ref{fig:sma-connector}).  
Two of such fibres were twisted around each other and the signal cable was soldered 
directly to the SMA connectors, whereas the other is grounded. 

%\begin{figure}[h]
%	\begin{center}
%		\includegraphics[width=8cm]{{pictures/DSCN1159.jpg}}
%		\caption{}
%		\label{fig:kabelhuelse}
%	\end{center}
%\end{figure}
%
%
%\begin{figure}[h]
%	\begin{center}
%		\includegraphics[width=8cm]{{pictures/DSCN1164.jpg}}
%		\caption{}
%		\label{fig:voreinzug}
%	\end{center}
%\end{figure}
%
%\begin{figure}[h]
%	\begin{center}
%		\includegraphics[width=8cm]{{pictures/DSCN1165.jpg}}
%		\caption{}
%		\label{fig:durchgezogen}
%	\end{center}
%\end{figure}
%
%\begin{figure}[h]
%	\begin{center}
%		\includegraphics[width=8cm]{{pictures/DSCN1153.jpg}}
%		\caption{}
%		\label{fig:sma-connector}
%	\end{center}
%\end{figure}

%\bigskip
%\begin{figure}[h]
%	\begin{center}
%		\includegraphics[width=8cm]{{pictures/crimp.png}}
%		\caption{Picture of cable with crimp}
%		\label{fig:crimp}
%	\end{center}
%\end{figure}

With this fabrication procedure some 20 cables have been produced with various lengths up to \SI{3}{\meter}. 
The procedure leads to a high efficiency and reproducibility, as will be shown in the next section.

\section{First Cable Performance Tests}
\label{sec:firsttests}
The principle functionality of the carbon cables was confirmed with some initial tests without the goal to find the maximal transmission frequency. The cable was inserted into a setup to read-out an actual front end module as used in the recently installed IBL\footnote{IBL: Insertable B-Layer} pixel detector~\cite{bib-IBLTDR},\cite{bib-IBL}, signal shape and propagation delay have been measured, and Bit-Error-Rate tests were performed using a commercial tester up to \SI{200}{\mega\hertz}. 
Furthermore, it was tested if the production of different cables leads to reproducible results.
These tests will be shortly described in the next subsections. 

\subsection{Controlling a Detector Module using a Carbon Cable}

\begin{figure}[h]
	\begin{center}
		\includegraphics[width=12cm]{{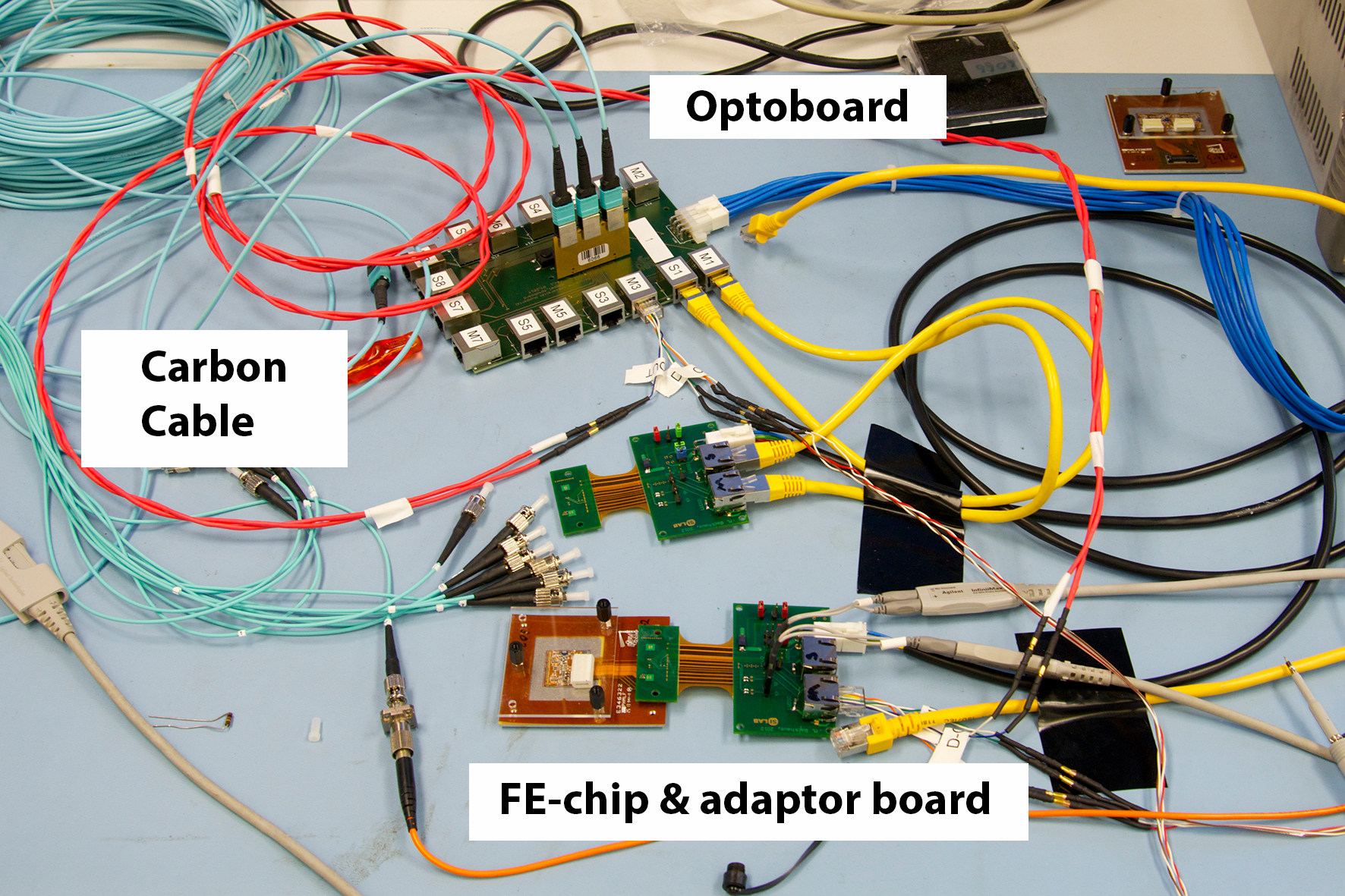}}
		\caption{Setup using carbon cables to connect a FE-chip to an opto-transceiver.}
		\label{fig:fe-setup}
	\end{center}
\end{figure}

A system, according to the one used in the actual detector, has been set up with all components used in the ATLAS pixel detector for testing and developing its read-out system \cite{bib-IBLreadout}. It contained the whole read-out chain from an up-to-date detector module to a read-out unit comprised of an FPGA-based steering board (IBL BOC-card). The BOC-card was connected optically to the opto-transceiver, which then connected the module via three electrical twisted pair cables. Clock and data to the module as well as data from the module were sent via these lines. Differential transmission was used on these cables, normally made from aluminium or copper. To test the transmission via a carbon cable, one of the three twisted pair wires was replaced by a carbon twisted pair cable at a time.   

First, the clock and steering data transmission running at \SI[per-mode=symbol]{40}{\mega\bit\per\second} was tested and the detector module could be operated with these signals running over carbon cables without any problem. The transmission speed via the carbon cable was finally increased up to \SI[per-mode=symbol]{160}{\mega\bit\per\second} by using the carbon cable on the data line from the module to the opto-transceiver. Also this configuration worked without problems. 

For this setup a series of operational tests have been performed. Normally the module continuously sends IDLE words, a bit series being transmitted when there is no data to be sent. These IDLE words are checked on the read-out unit side to synchronize the system. This worked without any issues and no de-synchronization was observed. Furthermore, the front-end module data path needs to be initiated via a trigger. To check for a normal operation a series of triggers were sent to the detector module, which returned an acknowledge signal via the carbon cable. All this data has been detected and checked in the readout unit without finding any errors. 

This first test indicated that the carbon cable could be used for detector operation at low bandwidth, so the next steps were to check for a higher data rate and the reliability of transmission.

\subsection{Pulse shape and signal transmission delay}

\begin{figure}[h]
	\begin{center}
		\includegraphics[width=12cm]{{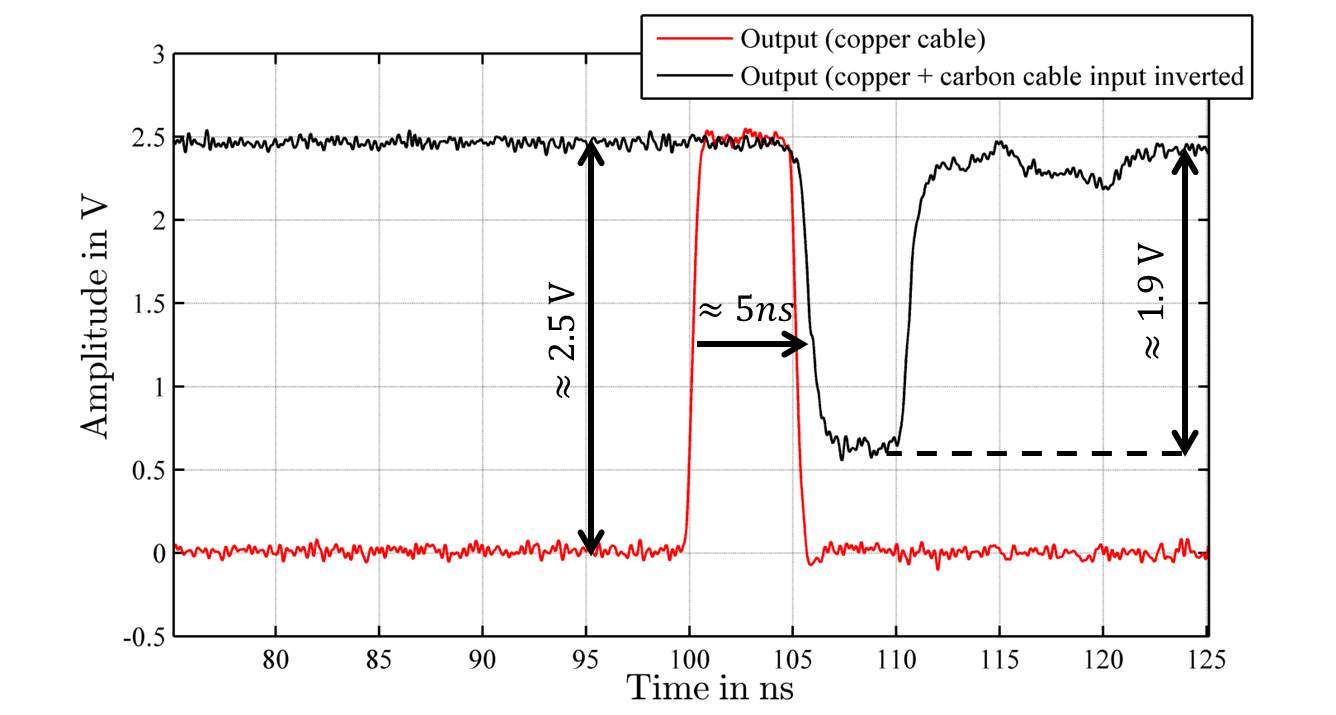}}
		\caption{Pulse delay and pulse shape for a \SI{1}{\meter} carbon cable}
	\label{fig:pulse_delay_shape}
	\end{center}
\end{figure}

To get an idea of the cable characteristics (i.e. pulse shaping, attenuation and delay) a pulse generator and an oscilloscope have been used. Since the pulse generator features two outputs (inverted and non-inverted) we can compare two transmission paths (carbon cable + copper cable and copper cable only) simultaneously. We measured a \SI{1}{\meter} carbon cable with a signal being sent with an original amplitude of \SI{2.5}{\volt}. Figure \ref{fig:pulse_delay_shape} shows a pulse delay of approximately \SI{5}{\nano\second} and the smoothed resulting pulse shape with a reduced maximal pulse amplitude of \SI{1.9}{\volt} (or an attenuation of \SI{-2.4}{\decibel}). Starting at \SI{115}{\nano\second} we can observe a weak reflected signal component indicating a mismatch of the characteristic impedance of the carbon cable and the input impedance of the oscilloscope (i.e. \SI{50}{\ohm}) 

\vspace{10pt}

\subsection{Bit Error Rate Measurements}

In a second test a standardized transmission test was performed using a commercial Bit Error Rate Tester (BERT). The device, Anritsu\textsuperscript{\textregistered} M1630B, is capable of testing %16 cable 
links with an internally or externally sent reference clock up to \SI{200}{\mega\hertz}.  
The BERT device sends pseudo random bit sequences (PRBS) based on variable lengths of the used shift register (7 to 31 bits)\footnote{A shift register length of $n$ results in a PRBS pattern length of $2^n-1$} to deliver a signal with an equal number of ones and zeros and providing every possible combination of bits for the given pattern lengths. 
At the receiving end the device registers incoming data and compares it to the pattern sequence sent out. The signal amplitude can be chosen from several logic level standards or manually set. For the carbon fibre tests a LV-TTL\footnote{Low Voltage Transistor-Transistor Logic, a standard logic level with \SI{3.3}{\volt} amplitude.} signal with an amplitude of \SI{3.3}{\volt} has been chosen at a duty cycle of \SI{50}{\percent}.

For the measurement different shift register lengths for the PRBS generation have been tested without a visible difference, so it was decided to use a pseudo random bit sequence based on register length of 11 bits for the measurements. 

To measure the error rate for a given cable, first it is mandatory to determine the phase shift and the sampling threshold for the receiving part. This ensures that the signal is synchronized correctly and the sampling threshold fits the incoming signal. 
Within 24~hours some \SI[per-mode=symbol]{17}{\tera\bit} were transmitted. No error was 
observed, which implies that the theoretically achievable bit error rate would be better than $BER=$\SI{5.8e-13}.
The bit error rate in this paper is simply derived by: 
\begin{equation}
BER = \frac{1}{\textrm{transmitted bits}}
\end{equation}
To state the bit error rate to a confidence level of \SI{95}{\percent}, about three times as many bits had to be sent.

% \begin{table}[ht]
%	\centering
%	\begin{tabular}{l l l l l}
%		Cable type & Cable length & Transmitted bits & Measured errors & Bit error rate \cr
%		\hline
%		\hline
%		&&&&\cr
%		\hline
%	\end{tabular}
%	\caption{Result of the measure bit error rates with the Anritsu BERT at 200~MHz.}
%	\label{tab:BERT}
%\end{table}

%\missingfigure{BERT schematic /  Setup}

\subsection{Reproducibility of Cables}

To verify if the fabrication process of the carbon cables is reproducible, four identical CF cables of one meter length were produced with the procedure described above. All of the cables passed the first tests for fractures or short circuits. 

In a next step the bit errors were tested over a period of at least 24 hours using the Anritsu\textsuperscript{\textregistered} BERT. No errors were observed in the some \SI[per-mode=symbol]{17}{\tera\bit} sent for either of these cables. 
% (see Tab~\ref{tab:repro}).

In a final step the S-parameters of the cables were measured and compared using a Rohde $\&$ Schwarz\textsuperscript{\textregistered} ZND 2-port network analyser \cite{NVA}. S-parameters offer an easy to measure characteristics of electronic devices with a certain number of ports (in our case we have two ports corresponding to the input and output port of the cable). The S-parameters characterize the frequency dependent behaviour of the cable while the signal input and the measurement output are interchanged. Therefore, the S-parameter $S_{11}$ describes the output measured at port 1 while an input signal is applied also to port 1. This can be interpreted as the reflection factor of the cable. The parameter $S_{21}$ describes the output observed at port 2 while an input signal is applied to port 1. Therefore, this parameter gives us some information about the signal transmission properties or filter characteristic of the cable. S-parameters measured with a network analyser are in general complex and thus give additional information about the phase shift of the signal applied to the cable. This information will be used later to calculate the time delay and the propagation speed on the carbon cable.

\begin{figure}[t]
	\begin{center}
		\includegraphics[width=12cm]{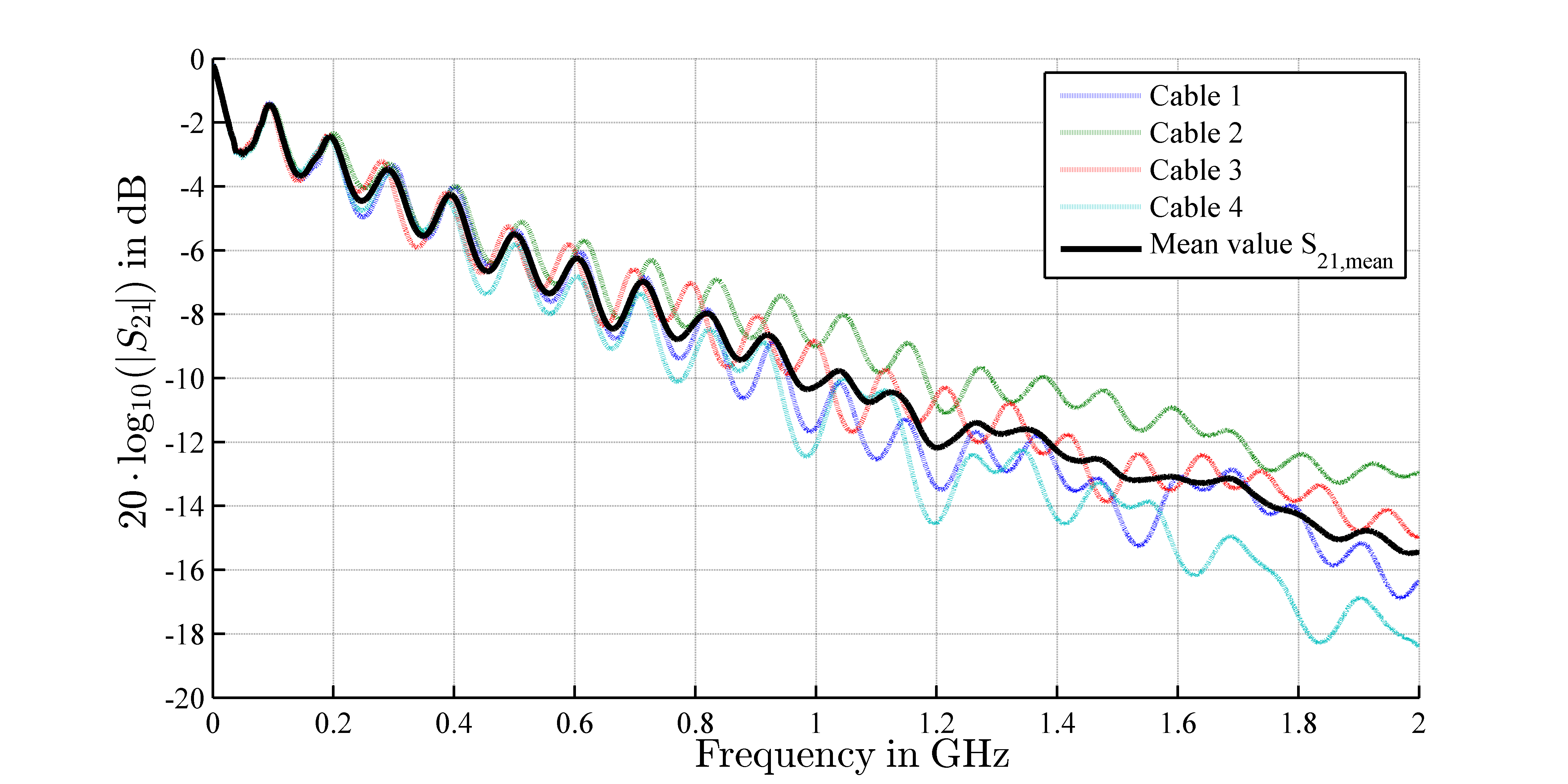}
		\caption{S-parameter $S_{21}^n$ for the four cables to test reproducibility of production process.}
	\label{fig:s21-parameters-consistency}
	\end{center}
\end{figure}

All  S-parameters were found to be consistent for the four cables. The absolute values of the $S_{21}$ parameters are shown together with the mean value in Figure~\ref{fig:s21-parameters-consistency}. The mean value $S_{21\text{,mean}}$ of the parameter $S_{21}$ is determined using 

\begin{equation}
S_{21\text{,mean}} = \frac{1}{4}~\sum_{n=1}^{4} |S_{21}^n| 
\end{equation}

(where $S_{21}^n$ corresponds to the $S_{21}$-parameter of cable $n$) and shown in Figure~\ref{fig:s21-parameters-consistency}. Parameters of the manufactured cables are in good agreement up to \SI{800}{\mega\hertz} with increasing difference for higher frequencies up to \SI{2}{\giga\hertz}. In our measurement cable 4 shows the lowest performance compared to the other three cables. If we assume that we have homogeneous carbon cables any deviations result
from slight differences during soldering and crimping of the connectors. In general, it can be observed that the attenuation of all cables increases for higher frequencies but stays well below \SI{-20}{\decibel} up to \SI{2}{\giga\hertz}. The strong oscillation of the absolute values of $S_{21}^n$ as well as of $S_{21\text{,mean}}$ results from standing waves which exist because of the mismatching of the characteristic impedance of the carbon cable and the $50\;\Omega$ impedance of the network analyser. 

To quantify the reproducibility of the cable manufacturing process all $S_{21}^n$ parameters are compared to $S_{21\text{,mean}}$ and the maximal difference $S_{21\text{,maxErr}}$ is determined using 

\begin{equation}
	S_{21\text{,maxErr}}=20\cdot \log_{10} (\max_{n=1\cdots 4}{(|S_{21}^n|-S_{21,\text{mean}}))}
\end{equation}
	\label{eq:maxdev}

\begin{figure}[t]
	\begin{center}
		\includegraphics[width=12cm]{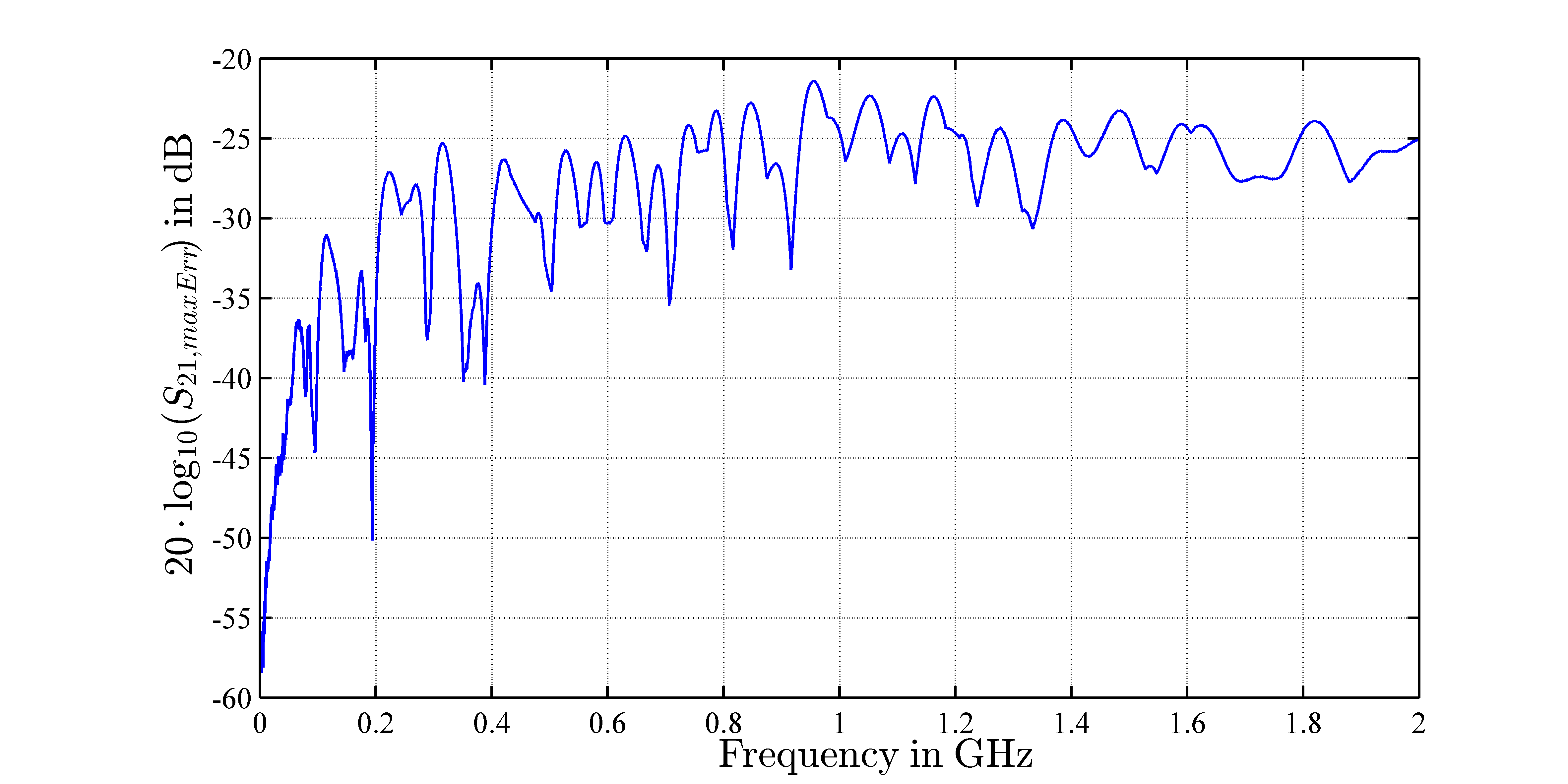}
		\caption{Maximal difference of the measured S-parameter $S_{21}$ compared to mean value. }
%		(See Eq.~\ref{eq:maxdev}).  }
		\label{fig:s21-error-deviation}
	\end{center}
\end{figure}

This difference can be seen in Figure~\ref{fig:s21-error-deviation}. $S_{21\text{,maxErr}}$ starts with a very small difference between the cables of approximately \SI{-55}{\decibel} for frequencies in the \SI{}{\kilo\hertz}-range. For higher frequencies in the GHz-range the difference increases but stays below -20~dB. Nevertheless, it is assumed that this value can be further reduced using some automated based production process.

\begin{figure}[t]
	\begin{center}
		\includegraphics[width=12cm]{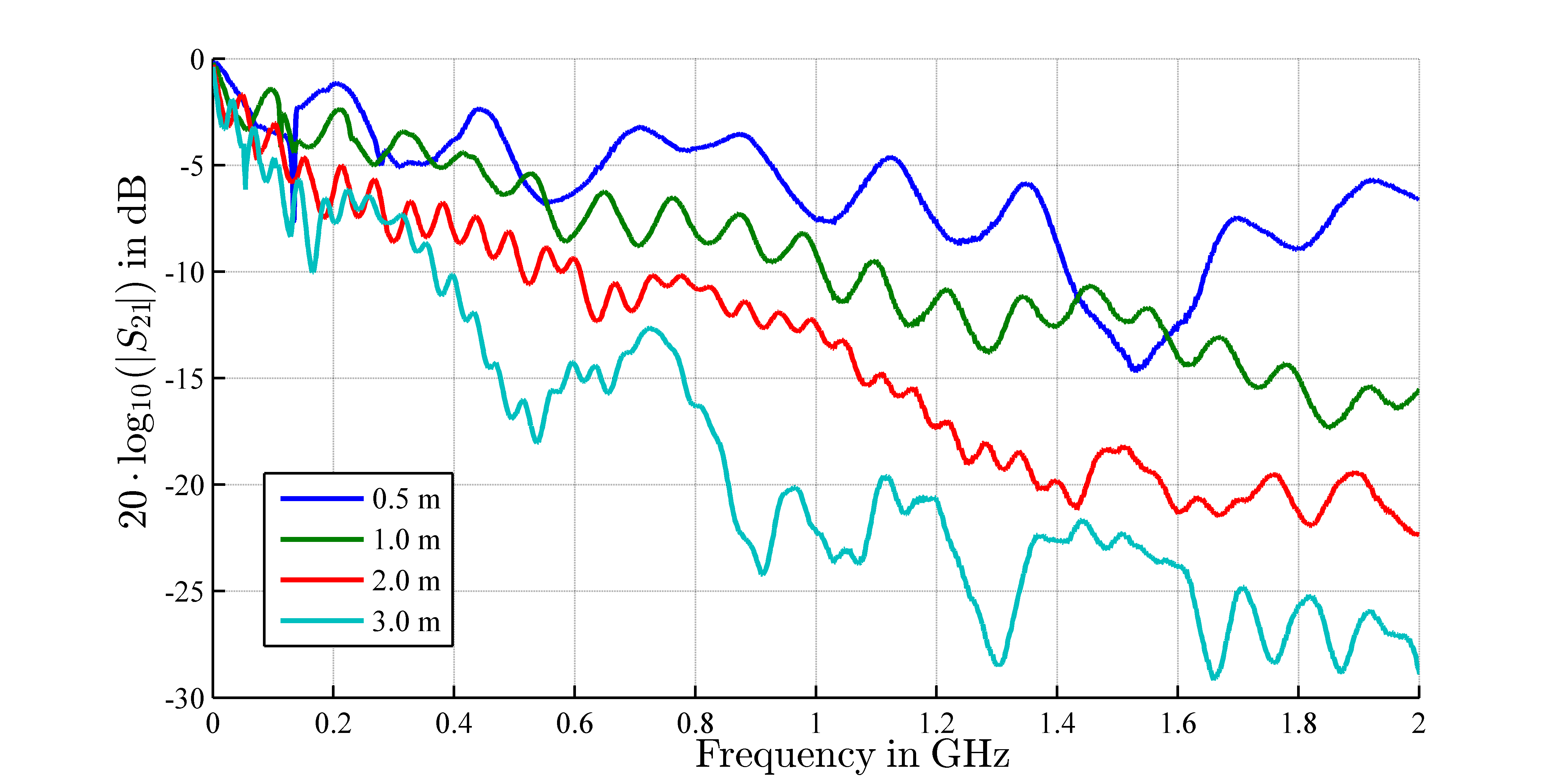}
		\caption{Measured S-parameter $S_{21}$ for different cable length.}
		\label{fig:s21-cab_len}
	\end{center}
\end{figure}

All measurements so far show that the transmission parameter $S_{21}$ for 1 m carbon cable is strongly frequency dependent. To receive a first impression of the effect of various cable lengths Figure~\ref{fig:s21-cab_len} compares $20\log_{10}(|S_{21}|)$ for cables of \SI{0.5}{\meter}, \SI{1}{\meter}, \SI{2}{\meter} and \SI{3}{\meter}.

\vspace{70pt}

Two important effects can be observed:
\begin{itemize}
	\item The maximal attenuation observed at \SI{1.2}{\giga\hertz} increases with the cable length from 7dB (\SI{0.5}{\meter} cable length) up to $\approx$ 25dB (for \SI{3}{\meter} cable length).
	\item The period of the oscillation observed for $S_{21}$ which results from the standing wave on the transmission line, depends on the length of the cable. 
\end{itemize}

% \vspace{10pt}
% \begin{table}[ht]
%	\centering
%	\begin{tabular}{l c c c c}
%		Cable & time [h] & Error Rate & Error Count & Frame Errors \cr
%		\hline
%		\hline
%		Cable\_1\_2 & 24 & $< 10^{-13}$ & 0 & 0 \cr
%		Cable\_3\_4 & 24 & $< 10^{-13}$ & 0 & 0 \cr
%		Cable\_5\_6 & 41 & $< 10^{-13}$ & 0 & 0 \cr
%		Cable\_7\_8 & 70 & $< 10^{-13}$ & 0 & 0 \cr
%		Cable\_9\_10 & 24 & $< 10^{-13}$ & 0 & 0 \cr
%		\hline
%	\end{tabular}
%	\caption{Bit error rate measurements of 5 identical 1~m long carbon cables
%	              {\bf (more precise: 72 has smaller BER than 24 h)}.}
%	\label{tab:repro}
% \end{table}

\section{Carbon Cable Characterization}
\label{sec:modell}
After describing the production of carbon cables and first measurements this section deals with the theoretical analysis of carbon fibres used as high speed data transmission line. 
We use similar approaches already known from transmission line theory for two wire twisted pair or coaxial transmission lines \cite{tl_model:pozar}. After a short introduction of the lumped element model we show how S-parameters, can be used to generate a model of the carbon fibre cable. Afterwards, this model can be used to perform simulations in time and frequency domain as well as investigate matching issues of the carbon cable using the standard circuit simulator Spice \cite{tl_model:spice}. Here we show that a model can be devised which agrees well with the data and can serve for future optimization. 

\subsection{Transmission line model}
\begin{figure}[t]
	\begin{center}
		\includegraphics[width=10cm]{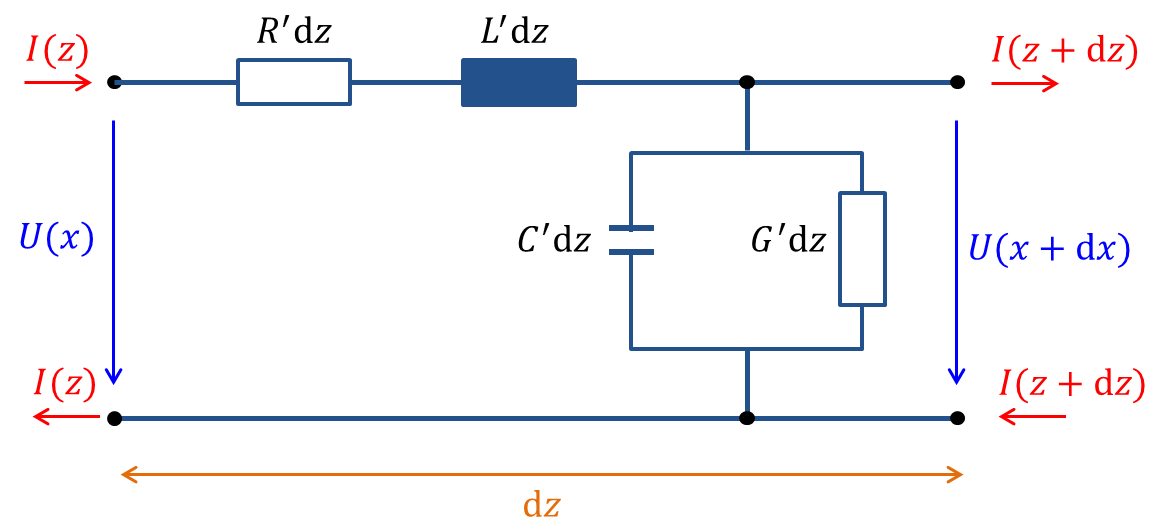}
		\caption{Transmission line model}
		\label{fig:tl_model}
	\end{center}
\end{figure}

As for standard copper cables (e.g. coaxial or twisted-pair) the classical transmission line model \cite{tl_model:pozar} is used as starting point. Figure~\ref{fig:tl_model} shows the general transmission line model consisting of lumped elements ($R^\prime$,$L^\prime$,$G^\prime$,$C^\prime$) which describe resistance, inductance, admittance and capacitance of a short piece of transmission line normalized to the length d$z$ of this line. It has to be considered that these elements are frequency dependent. A qualitative summary of these dependencies is shown in Table~\ref{tab:tl_param}. The final goal of our research is to quantify the frequency dependent values of the lumped elements in the transmission line model. Nevertheless, in this paper the primary goal is to develop a consistent flow to characterize the manufactured cables. Thus, we show the consistency of S-parameter measurements, time domain Spice simulations and real-time BER-tests.

\bigskip
			
\begin{table}[b]
	\caption{Frequency dependency of the transmission line parameters}
	\label{tab:tl_param}
	\smallskip
	\centering
	\begin{tabular}{lll}
		Parameter & Frequency dependency & Physical reason \\
		\hline
		\hline
		$R'$ & strongly increasing & Skin Effect	\\
		$L'$ & slightly reducing   & Skin Effect  \\
		$C'$ & nearly constant     & --						\\
		$G'$ & strongly increasing & Dielectrical Losses  \\
		\hline
	\end{tabular}	
\end{table}
			
As a first important parameter the group delay $\tau_{gd}$ of the carbon fibre cable can be determined based on the S-parameter measurements as
 
\begin{equation}
\tau_{gd}=-{d\phi}/{d\omega} \ \ \ \mathrm{with} \ \ \ \phi=\arctan (\Re(S_{21}) / \Im(S_{21}))
\end{equation} 

\begin{figure}[b]
	\begin{center}
		\includegraphics[width=12cm]{{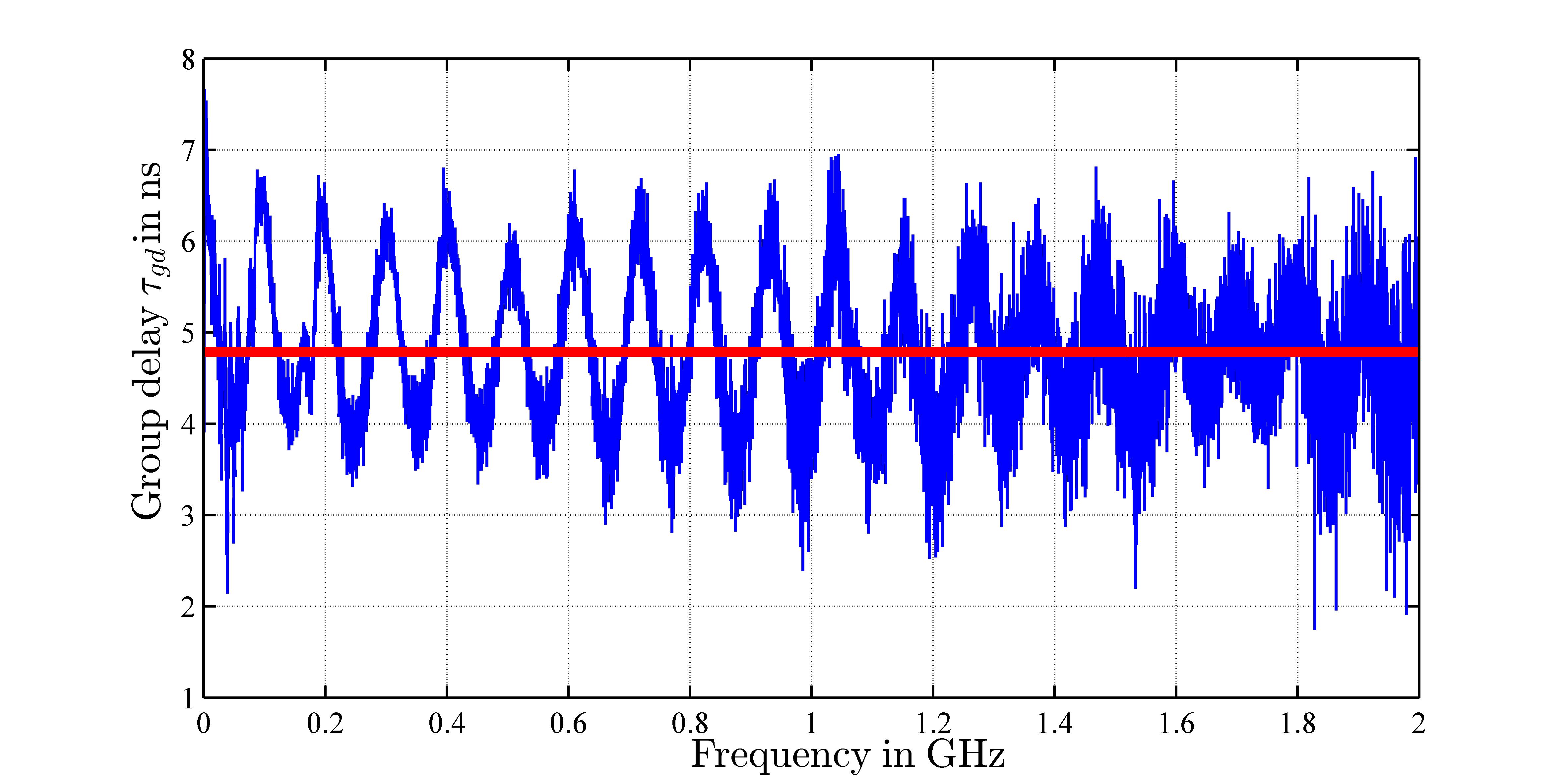}}
		\caption{Frequency dependent group delay $\tau_{gd}$ (blue) and mean value (red) of the carbon fiber, derived from measured $S_{21}$-parameter.}
		\label{fig:group_delay}
	\end{center}
\end{figure}

(here $\omega = 2 \pi f$). Figure~\ref{fig:group_delay} shows the frequency dependent $\tau_{gd}$ in blue and the mean value for frequencies $f=$\SI{100}{\kilo\hertz} ... \SI{2}{\giga\hertz} (red). In this plot the frequency dependency of the group delay can be clearly identified. Nevertheless, the mean value gives a group velocity of $v_{g} \approx$ \SI[per-mode=symbol]{5}{\nano\second\per\meter}, consistent with the result of Figure~\ref{fig:pulse_delay_shape}.
Based on the group velocity $v_g$ the so called velocity factor (\textit{VF}) can be determined using 

\begin{equation}
VF \ = \ \frac{v_g}{c}\ .
\end{equation}

For the twisted-pair carbon fibre cable discussed in this paper we receive \textit{VF}$=0.6867$ (here, it is assumed that the \textit{VF} is frequency independent).

\bigskip

\subsection{S-parameter measurements and broadband cable model}
In a next step the goal is to find the elements of an equivalent circuit which models the carbon fibre cable in a best possible way to receive a so called broadband model of the transmission line (i.e. a series connection of properly dimensioned short transmission line elements as shown in Figure~\ref{fig:tl_model}). 

\begin{center}
	\begin{figure}[t]
		\begin{center}
			\includegraphics[width=\textwidth]{{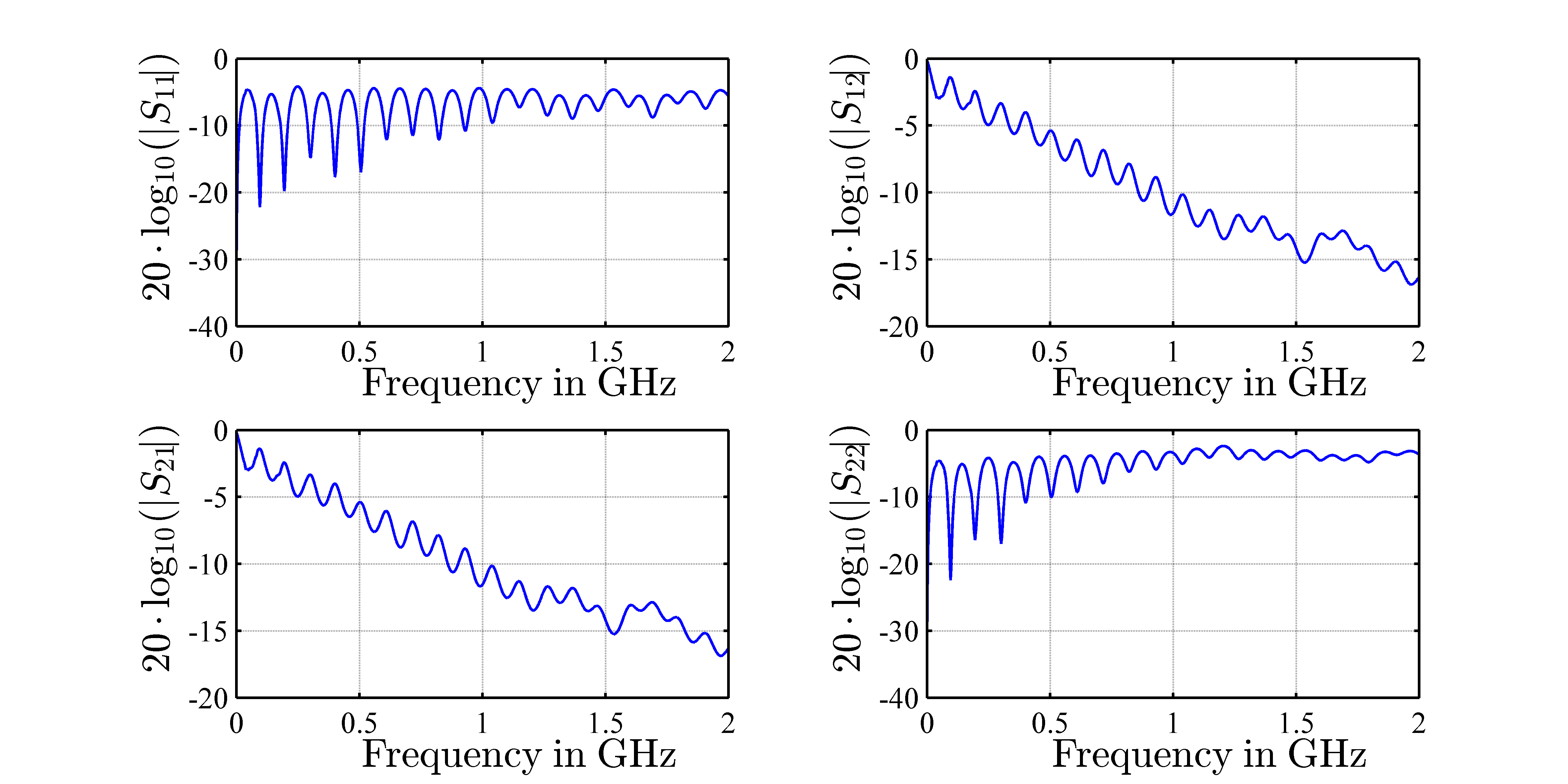}}
			\caption{Measured S-parameters of (\SI{1}{\meter} carbon fibre; Cable 1 in Figure~\protect\ref{fig:s21-parameters-consistency})} 
			\label{fig:s_param_meas}
		\end{center}
	\end{figure}
\end{center}

The four S-parameters (i.e. $S_{11}$, $S_{12}$, $S_{21}$, $S_{22}$) are shown for the \SI{1}{\meter} cable in Figure~\ref{fig:s_param_meas}. It can be seen that the two transmission parameters (i.e. $S_{12}$, $S_{21}$) are nearly symmetric and feature a maximum attenuation of approximately \SI{-17}{\decibel} at \SI{2}{\giga\hertz}. From a theoretical perspective the only difference between $S_{12}$ and $S_{21}$ is the direction in which the cable is used. From a practical perspective the slightly different crimping and soldering of the SMA-connectors lead to slightly different measurement results. For these first investigations we will neglect these differences.
Additionally, strong reflections leading to periodic dips for parameters $S_{11}$ and $S_{22}$ are observed,
which point to a mismatching of the twisted-pair and the coaxial cables. These reflecting components and the associated matching issues will be subject of future investigations.

\begin{center}
	\begin{figure}[h]
		\begin{center}
			\includegraphics[width=10cm]{{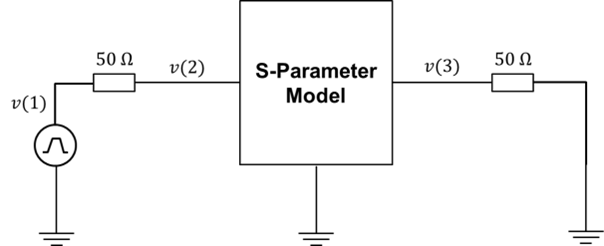}}
			\caption{S-Parameter simulation setup}
			\label{fig:sim_setup}
		\end{center}
	\end{figure}
\end{center}

The S-parameters are converted to a broadband Spice model using a commercial tool \cite{bb_model:ads} to perform circuit simulations in Spice and assess the characteristics of the transmission line and additional external components (i.e. load and inner impedance) during simulation in detail. This step is also required during optimization of the cable and possible impedance matching. The simulation setup with a pulse generator featuring a \SI{50}{\ohm} inner impedance and a \SI{50}{\ohm} load is shown in Figure~\ref{fig:sim_setup}. Moreover, simulations can help to find optimal cable properties and thus help to find optimal cable materials and structures.

As already mentioned, the first investigations concerning broadband model generation out of S-parameters were based on the commercial tool. Additionally, we compare the results of this tool \cite{bb_model:ads} and a freely available software (s2spice \cite{spice:s2spice}). Both tools show comparable results. 

\bigskip

\subsection{Spice simulation}
This subsection describes setting up a simulation environment for the carbon fibre cables. Spice simulations will allow us to study the time-domain behaviour of the transmission line. Here, the two aspects pulse shape and pulse delay are of major interest.

\begin{center}
	\begin{figure}[h]
		\begin{center}
			\includegraphics[width=10cm]{{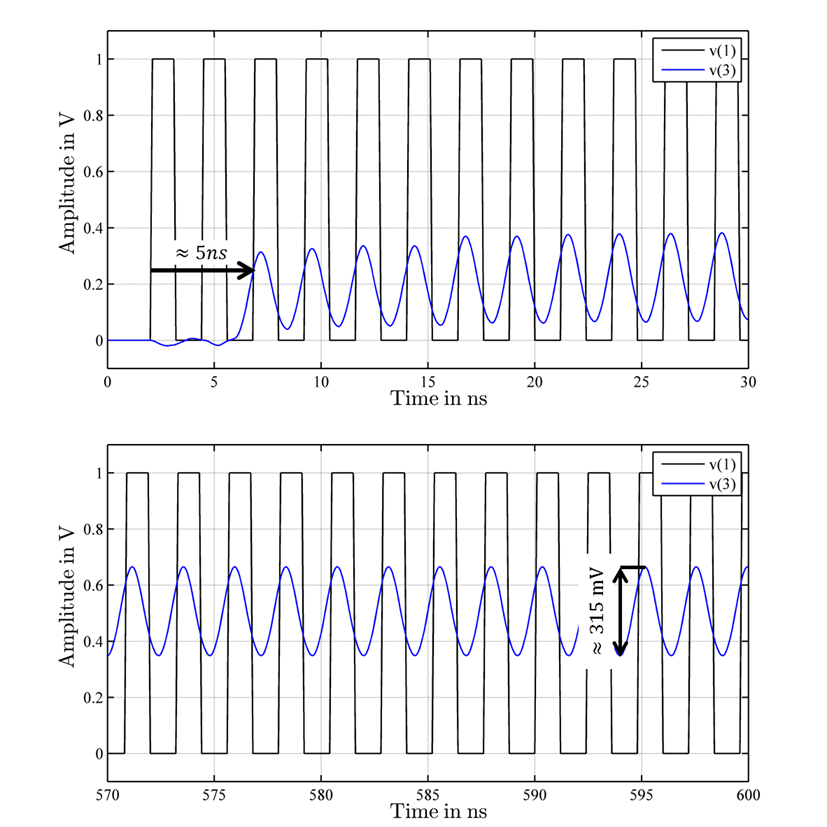}}
			\caption{Spice based broadband simulation of a 1~m carbon-cable. Input signal $v(1)$ (black) output signal $v(3)$ (blue). Upper part shows the beginning of the transmission while the lower part shows the steady state }
			\label{fig:delay_meas}
		\end{center}
	\end{figure}
\end{center}

In this section we show the consistency of the measured results and the predicted results using the simulation and the transmission line model to prove our approach. 
We use a Spice simulation tool freely available (e.g. LTSpice \cite{spice:ltspice}). A pulse train with adjustable pulse length (i.e. different data rates) is used as input signal during simulation. The results are shown in
Figure~\ref{fig:delay_meas} for an input signal with a data rate of \SI{1}{\giga\hertz}. The input pulse has an arbitrary peak voltage of \SI{1}{\volt}. The upper part of the figure shows the start of the transmission while the lower part shows the stable state of transmission with an expected offset value of 0.5 V.

\bigskip

Again, the delay (visible in the lower part of the figure) of approximately \SI[per-mode=symbol]{5}{\nano\second\per\meter} equals the measurements. It can be seen in the upper part of the figure that the output signal's amplitude is reduced to approximately one third of the input amplitude. This corresponds to \SI{10}{\decibel} attenuation determined during S-parameter measurement at 1 GHz.  Additionally, a filter effect can be observed (i.e. the cable causes a smooth shape of the output signal). This is caused mainly by the fact that the transmission line model acts as a low-pass filter (i.e. high frequencies of the input signal are significantly attenuated while low frequencies pass without attenuation). This effect can also be seen in the time domain Spice simulation and the time domain measurements using the oscilloscope described earlier. The low pass characteristic of the transmission line can be already seen in the equivalent circuit (see Figure~\ref{fig:tl_model}).

% For low frequencies the capacitor can be removed and the inductor can be replaced by a short-cut. For high frequencies the situation is vice versa. 
%The spice simulation will further allow us {\bf (but not done??)} to investigate the matching of the characteristic impedance of the transmission line and the load impedance and to simulate  
%eye diagrams and to investigate the optimal threshold for the binary decision in the receiver.

\section{FPGA based Bit Error Rate Test}
\label{sec:fastfpga}
\begin{center}
	\begin{figure}[h]
		\begin{center}
			\includegraphics[width=8cm]{{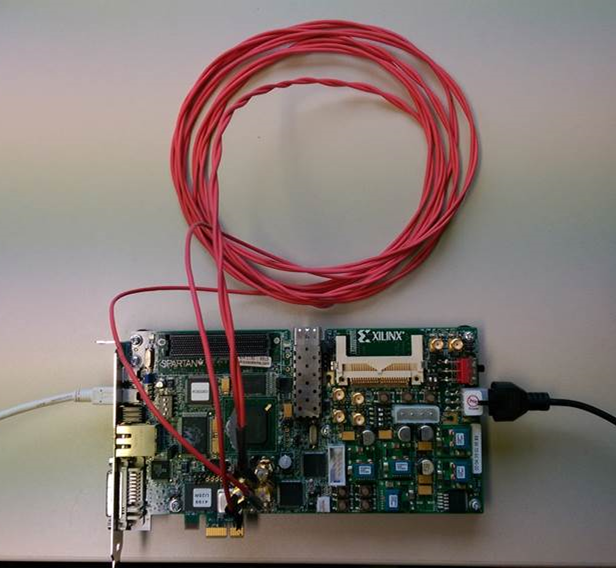}}
			\caption{FPGA-based cable BER-tester (cable is coiled up just for the photograph)}
			\label{fig:ber_tester}
		\end{center}
	\end{figure}
\end{center}

Measurements using the commercial BER-tester were limited to \SI{200}{\mega\hertz}. 
Nevertheless, investigations of the S-parameters and simulations suggest that higher transmission rates beyond \SI{1}{\giga\hertz} should be possible considering the observed attenuation. 
Therefore, as a final step the cable was verified using a BER-tester realized using a FPGA development board \cite{sp605:xilinx}. Figure~\ref{fig:ber_tester} shows the development board with the carbon fibre cable attached using differential signal transmission. The BER-tester software comes as VHDL-core alongside the FPGA integrated development environment \cite{IBERT_core:xilinx}. With the current FPGA-board setup we are restricted to certain frequencies commonly used by serial interface protocols. 

\bigskip

Various carbon fibre cables with different lengths have been tested at different transmission rates.
Results and associated BER values are shown in Table~\ref{tab:cab_ber}. During the FPGA-based BER-tests we used RX-equalization supported by the transceivers (parameter \textit{RxEq}). This feature allows one
to amplify higher frequencies in the transmitter which is required to compensate for the low pass characteristic of the cables, which have been identified during S-parameter measurements. Based on the current analysis flow these problems can be identified directly after the manufacturing process of the cables. The parameters \textit{TxEmp} controls the transmit voltage swing of the input signal (i.e. the peak-to-peak voltage of the input pulse). 

\begin{table}[htbp]
	\caption{Cable length vs. BER (transmission of \SI[per-mode=symbol]{10}{\tera\bit})}
	\label{tab:cab_ber}
	\smallskip
	\centering
	\begin{tabular}{lllll}
		Cable & Tx Rate & BER & TxEmp & RxEq \\
		\hline
		\hline
		\SI{0.5}{\meter} & \SI[per-mode=symbol]{0.625}{\giga\bit\per\second} & \num{1e-13}	 & \SI{0}{\decibel}  & \SI{0}{\decibel}  \\
		\SI{1.0}{\meter} & \SI[per-mode=symbol]{0.625}{\giga\bit\per\second} & \num{1e-13}	 & \SI{0}{\decibel}  & \SI{0}{\decibel}  \\
		\SI{2.0}{\meter} & \SI[per-mode=symbol]{0.625}{\giga\bit\per\second} & \num{1e-13}	 & \SI{0}{\decibel}  & \SI{0}{\decibel}  \\
		\SI{3.0}{\meter} & \SI[per-mode=symbol]{0.625}{\giga\bit\per\second} & \num{1e-13}	 & \SI{0}{\decibel}  & \SI{0}{\decibel}  \\
		\SI{0.5}{\meter} & \SI[per-mode=symbol]{0.781}{\giga\bit\per\second} & \num{1e-13}	 & \SI{0}{\decibel}  & \SI{0}{\decibel}  \\
		\SI{1.0}{\meter} & \SI[per-mode=symbol]{0.781}{\giga\bit\per\second} & \num{1e-13}	 & \SI{0}{\decibel}  & \SI{0}{\decibel}  \\
		\SI{2.0}{\meter} & \SI[per-mode=symbol]{0.781}{\giga\bit\per\second} & \num{1e-13}	 & \SI{0}{\decibel}  & \SI{0}{\decibel}  \\
		\SI{3.0}{\meter} & \SI[per-mode=symbol]{0.781}{\giga\bit\per\second} & \num{3e-13}	 & \SI{0}{\decibel}  & \SI{2.6}{\decibel} \\
		\SI{0.5}{\meter} & \SI[per-mode=symbol]{1.25}{\giga\bit\per\second}  & \num{1e-13}	 & \SI{0}{\decibel}  & \SI{0}{\decibel}  \\
		\SI{1.0}{\meter} & \SI[per-mode=symbol]{1.25}{\giga\bit\per\second}  & \num{1e-13}	 & \SI{0}{\decibel}  & \SI{0}{\decibel}  \\
		\SI{2.0}{\meter} & \SI[per-mode=symbol]{1.25}{\giga\bit\per\second}  & \num{4e-13}	 & \SI{7.6}{\decibel} & \SI{2.6}{\decibel} \\
		\SI{3.0}{\meter} & \SI[per-mode=symbol]{1.25}{\giga\bit\per\second}  & \num{4.6e-12} & \SI{7.6}{\decibel} & \SI{2.6}{\decibel} \\
		\hline
	\end{tabular}	
\end{table}

Table~\ref{tab:cab_ber} shows that \SI[per-mode=symbol]{1.25}{\giga\bit\per\second}
can be transmitted over a carbon cable of up to \SI{2}{\meter} with an error rate of less than the
required \num{e-12} mentioned in the introduction. However, first errors occur at theattenuatorse
lengths and in general, as expected, the number of errors increases with length and
transmission rate. 
As a cross check we measured BER-values for a short (30 cm) copper-twisted-pair cable manufactured in the same way like the carbon fibre cables. During the BER-tests we receive no errors for \SI[per-mode=symbol]{10}{\tera\bit} of data transmitted at \SI[per-mode=symbol]{1.25}{\giga\bit\per\second}. Afterwards, we connect some attenuators in series to the copper-twisted-pair cables and receive comparable BER-values for  values of approximately \SI{15}{\decibel}.
This shows that the BER-values determined with the FPGA-based BERT are in good agreement with the S-parameter measurements (i.e. measured attenuation for $S_{21}$)

\section{Conclusions and Outlook}
\label{sec:Conclusions}
% \todo[inline, color=yellow]{Ausblick / weitere Anwendungen(?) (Hier sollten die wesentlichen Punkte zur Anpassung untergebracht werden. Außerdem könnte schon eine Abbildung zeigen, was passiert, wenn man den Netzwerkanalysator bis 6 GHz laufen lässt und damit das Frequenzspektrum der S-Parameter erweitert). Sollte das Altera FPGA noch kommen sollte man durchaus einen Vergleich der beiden Boards einfügen.}

In this paper we have discussed first studies if high bandwidths signals can be 
transmitted via carbon cables. Single wire cables made of 12000 nickel coated filaments
were read out. To reduce noise a second cable was twisted around the signal
cable. Although rather delicate, the fabrication process leads to reproducible results. Extraction of the Spice model and simulations in time and frequency domain are consistent with measurements. This proves the validity of the whole design process (see Figure~\ref{fig:prop_design_flow}).

Our first attempts lead to encouraging results. Without detailed optimization of the 
termination and choice of the fibre we already observe that \SI[per-mode=symbol]{1.25}{\giga\bit\per\second} can be transmitted 
over a length of about more than meter with an error rate of less than \num{e-12}. 
As expected from the transmission line model, increasing the cable length and the frequencies leads to an increase in the BER.

Whereas this paper reports on first studies to establish a proof of principle, further improvements and more systematic studies are planned. Given the good agreement of data and simulation, the 
model based on a frequency dependent lumped element description of carbon fibre and connector can be used for
further optimization. This model should also include any mismatching issues. Based on this model it will be possible to specify clear guidelines for the transmission line (i.e. fibre and sleeve material, number of filaments, thickness, connectors) as well as for the manufacturing process. 

% Another question is in how far the signal amplification in the transmitter used during FGPGA-based real-time BER-tests can be realized in the environment of the LHC upgrade. Other means could be the use of carbon fibres with less resistivity. 

Several measurements should be performed to systematically improve the performance and identify the potentials. E.g.
different types of filaments with and without coating and varying thickness will be systematically studied and characterized.
One of the aims is also to reduce the thickness, respectively radiation length of the cables and to optimize the connection.
Because of the large difference between the characteristic impedance of twisted-pair (balanced) realization of the carbon fibre cables and classical coaxial \SI{50}{\ohm} equipment (unbalanced) used in the presented test-setup and also in the final application, some broadband transformation from unbalanced to balanced and vice versa have to be investigated. These investigations should be based on the rather good description of the transmission line with a Spice model.

\acknowledgments

The authors would like to thank Prof. Dr.-Ing. Dirk Fischer and his team for support and discussion during S-parameter measurements, Camille Fausten for her support in measuring the S-parameters, as well as Peter Kind for detailed discussions on the topic.

\end{document}